\documentstyle[prd,aps,amsmath,amssymb]{revtex}

\begin{document}
\title{Relativistic Massive Vector Condensation}
\author{Francesco {\sc Sannino}\footnote{Electronic address: {\tt francesco.sannino@nbi.dk}}
\quad and \quad Wolfgang {\sc Sch\"afer} \footnote{Electronic
address: {\tt schaefer@nordita .dk}} }
\address{NORDITA, Blegdamsvej 17,
DK-2100 Copenhagen \O, Denmark}

\date{November 2001}
\maketitle

\begin{abstract}
At sufficiently high chemical potential massive relativistic spin
one fields condense. This phenomenon leads to the spontaneous
breaking of  rotational invariance while linking it to the
breaking of internal symmetries. We study the relevant features of
the phase transition and the properties of the generated Goldstone
excitations. The interplay between the internal symmetry of the
vector fields and their Lorentz properties is studied. We predict
that new phases set in when vectors condense in Quantum
Chromodynamics like theories.  \vskip 2cm
\end{abstract}


\section{Introduction}
\label{uno}

The Quantum chromodynamics (QCD) phase diagram as a function of
temperature, chemical potential and the number of light flavors is
very rich and highly structured. QCD should behave as a color
superconductor for a sufficiently large quark chemical potential
\cite{REV}. This leads to new phenomenological applications
associated with the description of quark stars, neutron star
interiors and the physics near the core of collapsing stars
\cite{REV,OS,HHS}. Much is known about the phase structure of QCD
at nonzero temperature through a combination of perturbation
theory and lattice simulations while the phase structure at
nonzero chemical potential and for large numbers of flavors has
been less extensively explored \cite{MariapaolaI}. Standard
importance sampling methods employed in lattice simulations fail
at nonzero chemical potential for $N_c=3$ since the fermionic
determinant is complex. {}For $N_c=2$, the situation is very
different since the quarks are in a pseudoreal representation of
the gauge group and lattice simulations can be performed
\cite{Morrison:1998ud,Hands:1999md,Aloisio:2000nr,Liu:2000in,Aloisio:2000if,Bittner:2001rf,Hands:2001hi,Muroya:2001qp,Hands:2001yh,Aloisio:2001rb,Kogut:2001if,Kogut:2001na,KST,KSTVZ,KT}.

At non zero chemical potential Lorentz invariance is explicitly
broken down to the rotational subgroup $SO(3)$ and higher spin
fields can condense, thus potentially enriching the phase diagram
structure of QCD and QCD-like theories. Indeed in
\cite{Lenaghan:2001sd} it has been suggested that some of the
lightest massive vectors at non zero baryon chemical potential may
condense. Recent lattice simulations seem to support these
predictions \cite{MPLombardo}. In the context of the Electroweak
theory  vector condensation in the presence of a strong external
magnetic field has also been studied in \cite{Ambjorn:1989gb}.

 In this paper we
explore the condensation of relativistic massive vectors when
introducing a non zero chemical potential via an effective
Lagrangian approach. In order for the vectors to couple to the
chemical potential they must transform under a given global
symmetry group. In view of the possible physical applications we
consider the vectors to belong to the adjoint representation of
the $SU(2)$ group.

To set the stage for the more complicated case of the vector field
we first consider a scalar field in the adjoint representation of
$SU(2)$ at non zero chemical potential. We choose the chemical
potential to lie along one of the generator's directions
explicitly breaking $SU(2)$ to $U(1)\times Z_2$. We then
investigate the dispersion relations of the 3 fields as function
of the chemical potential and show the existence of a second order
phase transition when the chemical potential equals the zero
density mass of the scalar field. In the broken phase we observe
the emergence of one gapless mode associated to the spontaneous
breaking of the $U(1)$ symmetry possessing linear dispersion
relations. This is consistent with the Chadha-Nielsen
\cite{Nielsen} counting rule for Goldstone's bosons when Lorentz
invariance is not assumed.

The situation becomes more involved when turning to the spin-one
massive field case. When the vector field condenses the rotational
$SO(3)$ invariance breaks spontaneously together with part of the
flavor (i.e. global) symmetry of the theory. The condensate locks
together space and flavor symmetries. {}Following the standard
mean-field approach we identify our order parameter with the
vector field itself.

In order to elucidate the mechanism we construct the simplest
effective Lagrangian at zero chemical potential. This consists of
a standard kinetic type term and a tree level potential
constructed as a series expansion in the order parameter. We
truncate our potential at the fourth order in the fields while
retaining all the terms allowed by Lorentz and $SU(2)$ symmetry.
The introduction of the potential term allows us to describe the
phase transition in some detail while our results will be general.

We prove that the number of gapless states emerging when the
flavor$\times$rotational symmetry breaks spontaneously is
insensitive to the choice of the coefficients in the effective
potential. However the momentum dependence of the associated
dispersion relations is heavily affected by the choice of the
potential. The latter is related to the differences between the
symmetries of the potential with respect to the ones of the
kinetic type terms. The full vector Lagrangian (i.e. kinetic $-$
potential) always possesses the $Z_2\times U(1)\times SO(3)$
(flavor$\times$rotational) symmetry which at high enough chemical
potential breaks spontaneously to $Z_2\times SO(2)$. The gapless
modes are linked to the 3 broken generators. One state is an
$SO(2)$ scalar while the other two states constitute a vector of
$SO(2)$. The scalar Goldstone possesses linear dispersion
relations independently of the choice of the vector potential and
it resembles the gapless mode of the scalar theory. However the
$SO(2)$ vector state can have either quadratic or linear
dispersion relations. In particular (as summarized in the Table I)
the quadratic ones occur when the potential parameters are such
that it possesses an enhanced $SO(6)$ symmetry. When condensing
the vector breaks the $SO(6)$ symmetry of the potential down to
$SO(5)$ while the kinetic-type term still has only a $U(1)\times
SO(3)$ invariance. In this case the potential has 5 flat
directions (i.e. 5 null curvatures) and we would count in the
spectrum 2-gapless vectors and an $SO(2)$ scalar. However the
reduced symmetry of the kinetic term prevents the emergence of two
independent gapless vectors while turning the linear dispersion
relations of one vector state into quadratic ones. This result is
also in agreement with the Chadha-Nielsen counting scheme. However
since the Goldstone modes with quadratic dispersion relations must
be counted twice relatively to the ones with linear dispersion
relations we saturate the number of broken generators associated
with the symmetries of just the potential term which is larger
than the ones associated with the symmetries of the full
Lagrangian. It is certainly important to investigate, in the
future, if the choice of the parameter space leading to an $SO(6)$
symmetry for the potential term is stable against quantum
corrections.

Our model can be related immediately to the physics of 2 colors
QCD which has been partially analyzed in \cite{Lenaghan:2001sd}
and is within reach of present lattice studies. Here we suggest
that vector condensation can be the leading mechanism for a
superfluid transition for 2 color QCD with one flavor. Indeed in
this case the global symmetry is $SU(2)$ which remains intact at
zero chemical potential and no Goldstone bosons emerge. The
massive spectrum of the theory still contains an $SU(2)$ spin-one
massive multiplet. When turning on a non zero chemical baryon
potential the vector condensation can lead to a novel superfluid
phase transition. Clearly our model can also be applied to QCD
with 3 color at non zero isospin chemical potential
\cite{Splittorff:2001mm}.

The paper is structured as follows. In section
\ref{sec:vec-and-mu} we first consider a toy model for a scalar
field in the regular representation of $SU(2)$ at non zero
chemical potential and investigate all of the relevant features of
the model. We then propose a simple effective Lagrangian for
vectors which allows us to uncover the main features related to
the condensation of relativistic massive spin-one fields. We
conclude and point out some physical applications in section
\ref{applications}. In particular we suggest that new phases
should set in for QCD like theories. Finally we extend our simple
model to $D-1$ number of space dimensions (with Euclidean $SO(D)$
Lorentz symmetry).

\section{Vector Condensation at Nonzero Chemical Potential}
\label{sec:vec-and-mu}

\subsection{Toy Model: Scalar Theory}

In order to illustrate and disentangle the main problems related
to the vector condensation we first consider the case of a scalar
field in the adjoint representation of an $SU(2)$  global
symmetry. This will allow us to show the feature which only
depends on the global symmetries first. The toy model Lagrangian
is:
\begin{equation}
{\cal L}={\rm Tr}\left[\partial_{\mu}\phi
\partial^{\mu}\phi\right]-m^2 {\rm Tr}
\left[\phi^2\right] - \lambda \left({\rm
Tr}\left[\phi^2\right]\right)^2 \ ,
\end{equation}
with $\phi=\phi^a T^a$ and $a=1,2,3$ and $T^a=\tau^a/2$. $\phi^a$
is a real field and $\tau^a$ the standard Pauli matrices.

We introduce the chemical potential associated with the $T^3$
generator. We shall see that our choice turns out to be
particularly relevant when considering 2 colors QCD and one flavor
at non zero quark chemical potential. The effects of the chemical
potential are included by generalizing the covariant derivative as
follow:
\begin{equation}
 D_{\mu}\phi = \partial_{\mu}\phi - i \mu \delta_{\mu
 0}\left[T^3,\phi\right]\ .
\end{equation}
$\mu$ is the chemical potential. Expanding the kinetic term we
have that the Lagrangians reads in components
$\phi=({\varphi^{\pm}}=(\phi^1 \mp i\,\phi^2)/\sqrt{2},~
\psi=\phi^3)$:
\begin{eqnarray}
{\cal L} &=& \frac{1}{2} (\partial \psi)^2 + |\partial \varphi|^2
+ \mu \,  \varphi^{(+)*} i\partial_0 \varphi^{(+)} - \mu \,
\varphi^{(-)*} i\partial_0 \varphi^{(-)} - \left(m^2 -
\mu^2\right) |\varphi|^2 - \frac{m^2}{2} \psi^2 -
\frac{\lambda}{4} \left(|\varphi|^2 + \psi^2 \right)^2 \, .
\end{eqnarray}
Our choice of the chemical potential explicitly breaks the $SU(2)$
'flavor'-symmetry to a $U(1)$ acting on the complex field $\phi$
and a discrete $Z_2$ symmetry for the neutral field $\psi$. {}For
$\mu \leq m$, the model has a unique vacuum at $|\varphi| = \psi =
0$ and the effect of the chemical potential is simply to split the
energies of the charged states as follows:
\begin{eqnarray}
E_{\varphi^\pm}(\vec{p}) =\mp \mu + \sqrt{m^2 + \vec{p}^2} \ ,
\qquad E_{\psi}(\vec{p}) = \sqrt{m^2 + \vec{p}^2} \ .
\end{eqnarray}
The mass-gaps are defined as the energy of each state evaluated at
zero momentum. Expanding the dispersion relations for small
momenta gives:
\begin{eqnarray}
E_{\varphi^{\pm}}(\vec{p}) = (m\mp\mu) + \frac{\vec{p}^2}{2m} +
{\cal O}(\vec{p}^4) \ . \label{nonrel}
\end{eqnarray}
This leads to the presence of a gapless excitation with a {\it
nonrelativistic} type of dispersion relation at $\mu = m$. Note
that due to the loss of Lorentz--covariance the mass-gaps do not
coincide with the zero density masses. The latter are defined as
the curvature of the potential evaluated on the vacuum and are
$m_{\varphi^{\pm}}^2 = m^2-\mu^2$, $m_\psi^2 = m^2$. Since the
$\psi$ field is not affected at all by the chemical potential it
does not condense and its mass-gap is $m_{\psi}=m$ for $\mu \leq
m$.

{}For $\mu > m$ the classical potential is unstable when $\phi=0$
and we choose the global minimum for
\begin{equation}\langle \varphi^1 \rangle
= \sqrt{\frac{\mu^2 - m^2}{\lambda}} \ ,
\end{equation}
breaking the $U(1)\times Z_2$ symmetry to $Z_2$. Computing the
curvature of the potential on the new vacuum for all the fields
yields:
\begin{eqnarray}
m^2_{\psi} = \mu^2  \ , \quad m^2_{\phi^1} = 2\left(\mu^2 - m^2\
\right) \ , \quad  m^2_{\phi^2} = 0 \ .
\end{eqnarray}
Note that in this phase the $\psi$ mass is given precisely by the
chemical potential. This is so since $\psi$ is not charged under
the chemical potential. {}After diagonalizing the quadratic terms
we obtain the following dispersion relations:
\begin{mathletters}
\begin{eqnarray}
E_{\pm}(\vec{p}) = \sqrt{ \vec{p}^2 + \bar{\mu}^2 \pm
\sqrt{\bar{\mu}^4  + 4 \vec{p}^2 \mu^2}} \, ,
\end{eqnarray}
\end{mathletters}
with $\bar{\mu}^2 = 3 \mu^2 - m^2$. Note that with our choice of
the vev $\varphi^{\pm}$ are no longer eigenstates. We see that a
gapless mode persists and the massive one has the mass gap
$\sqrt{2} \bar{\mu}$. At small momenta the dispersion relation of
the gapless (Goldstone-) mode is
\begin{mathletters}
\begin{eqnarray}
E_{-}^2(\vec{p}) \approx  {\mu^2 - m^2 \over \bar{\mu}^2} \,
|\vec{p}|^2
 + {2 \mu^4 \over \bar{\mu}^6} \,|\vec{p}|^4 \ .
\end{eqnarray}
\end{mathletters}
Hence in the broken phase we obtain the standard superfluid
Goldstone boson with an energy depending {\it{linearly}} on the
momentum, $E_{-} \sim v_s(\mu) \,|\vec{p}|$. However the {\it
velocity} $ v_s(\mu)$ vanishes at the transition point $v_s(\mu =
m) = 0$ turning the dispersion relation of the gapless excitation
into quadratic ones (i.e. $E_{-} \sim |\vec{p}|^2/2m$). This
behavior can be better understood by rewriting the velocity as
function of $m^2_{\phi^1}$ as follows:
\begin{equation}
v^2_s = {m^2_{\phi^1} \over m^2_{\phi^1} + 4 \mu^2} \ .
\end{equation}
We observe that the superfluid velocity depends on the curvature
of the potential in the direction orthogonal to the gapless
excitation. At the second order phase transition point this
curvature is zero, reflecting the conformal invariance of the
potential for the $\varphi$--fields. In the absence of Lorentz
breaking we have two massless modes at this point both with linear
dispersion relations. At nonzero chemical potential only one mode
is gapless while the information about the conformal invariance is
now inherited by the quadratic dispersion law. We shall see in the
next section, that this behavior is quite general, i.e. whenever
the potential has more symmetries than the derivative terms, this
information is encoded nontrivially into the momentum dependence
of the dispersion relations. Furthermore we can now smoothly
connect the dispersion relations of each state when traversing the
different regions of the phase diagram. Our result is in agreement
with the Chada-Nielsen \cite{Nielsen} counting scheme as well as
with recent studies related to the physics of color
superconductivity \cite{Miransky:2001tw,Schafer:2001bq}.


\subsection{The Vector Story}
There are different ways to introduce vector fields at the level
of the effective Lagrangian (for example the hidden local gauge
symmetry of Ref.\ \cite{BKY}, or the antisymmetric tensor field of
Ref.\ \cite{Ecker:1993de}) and they are all equivalent at
tree-level. We choose to introduce the massive vector fields
following the method outlined in \cite{KRS,KS,ARS,DRS}. This
method also allows a straightforward generalization of our model
to an arbitrary number of space dimensions. We adopt the following
model Lagrangian for describing relativistic spin one fields (in
$3+1$ dimensions) belonging to the regular representation of
$SU(2)$
\begin{eqnarray}
{\cal
L}&=&-\frac{1}{4}F^a_{\mu\nu}F^{a{\mu\nu}}+\frac{m^2}{2}A_{\mu}^a
A^{a\mu} -
 \frac{\lambda}{4}\left(A^a_{\mu}A^{a{\mu}}\right)^2 +
 \frac{\lambda^{\prime}}{4} \left(A^a_{\mu}A^{a{\nu}}\right)^2 \ ,
 \end{eqnarray}
with $F_{\mu
\nu}^a=\partial_{\mu}A^a_{\nu}-\partial_{\nu}A^a_{\mu}$, $a=1,2,3$
and metric convention $\eta^{\mu \nu}={\rm diag}(+,-,-,-)$. Here
$m^2$ is the tree level mass term and $\lambda$ and
$\lambda^{\prime}$ are positive dimensionless coefficients with
$\lambda > \lambda^{\prime}$. Other possible choices of the
parameters do not guarantee stability of the potential and will
not be considered in the following. This Lagrangian is not
renormalizible, contrary to the scalar theory presented above.
However it possesses the same global symmetries as the toy model.

The effect of a nonzero chemical potential associated to a given
conserved charge - related to the generator (say $B$) - can be
readily included \cite{Lenaghan:2001sd} by modifying the
derivatives acting on the vector fields:
\begin{equation}
\partial_{\nu} A_{\rho} \rightarrow \partial_{\nu}A_{\rho} - i
\left[B_{\nu}\ ,A_{\rho}\right]\ ,
\end{equation}
with $B_{\nu}=\mu \,\delta_{\nu 0} B\equiv V_{\nu} B$ where
$V=(\mu\ ,\vec{0})$. The vector kinetic term modifies according
to:
\begin{eqnarray}
{\rm Tr} \left[F_{\rho \nu} F^{\rho \nu}\right] &\rightarrow& {\rm
Tr} \left[F_{\rho \nu} F^{\rho \nu}\right] - 4i{\rm Tr}
\left[F_{\rho\nu}\left[B^{\rho},A^{\nu}\right]\right] - 2 {\rm Tr}
\left[\left[B_{\rho},A_{\nu}\right]\left[B^{\rho},A^{\nu}\right] -
\left[B_{\rho},A_{\nu}\right]\left[B^{\nu},A^{\rho}\right]\right]
\ .
\end{eqnarray}
The terms due to the kinetic term, after integration by parts,
yields \cite{Lenaghan:2001sd}
\begin{eqnarray}
{\cal L}_{kinetic}&=&\frac{1}{2}A_{\rho}^a\left\{ \delta_{ab}
\left[g^{\rho\nu}\Box-\partial^{\rho}\partial^{\nu}\right]
-4i\gamma_{ab}\left[g^{\rho \nu}V\cdot \partial - \frac{V^{\rho}
\partial^{\nu} + V^{\nu}
\partial^{\rho} }{2} \right] +  2
\chi_{ab}\left[V\cdot V g^{\rho\nu}-V^{\rho}V^{\nu}\right]\right\}
A_{\nu}^b
\end{eqnarray}
with
\begin{equation}
\gamma_{ab}={\rm Tr}\left[T^a\left[B,T^b\right] \right]\ , \qquad
\chi_{ab}={\rm
Tr}\left[\left[B,T^a\right]\left[B,T^b\right]\right] \ .
\end{equation}
{}For $B=T^3$ we have \begin{equation}
\gamma_{ab}=-\frac{i}{2}\epsilon^{ab3} \ , \quad
\chi_{11}=\chi_{22}=-\frac{1}{2} \ , \quad \chi_{33}=0 \ .
\end{equation} The chemical potential induces a ``magnetic-type''
mass term for the vectors at tree-level. The similarities with
respect to the scalar toy model as well as the symmetries of the
vect Lagrangian are more easily understood using the following
Euclidean notation:
\begin{equation}
\varphi_M^a=(A_M^1,A_M^2) \ , \qquad  \psi_M=A_M^3 \ ,
\end{equation}
with $A_M=(iA_0,\vec{A})$ and metric signature ($+,+,+,+$).  In
these variables the potential reads:
\begin{eqnarray}
V_{Vector}&=&\frac{m^2}{2}\left[|\vec{\varphi}_{0}|^2  + \psi_M^2
\right] +\frac{m^2-\mu^2}{2} |\vec{\varphi}_I|^2
+\frac{\lambda}{4}\left[|\vec{\varphi}_{M}|^2 + \psi_M^2\right]^2-
\frac{\lambda^{\prime}}{4}\left[\vec{\varphi}_{M}\cdot
\vec{\varphi}_{N} + \psi_M \psi_N\right]^2
\end{eqnarray}
with $I=X,Y,Z$ while $M,N=0,X,Y,Z$ and repeated indices are summed
over. At zero chemical potential $V_{Vector}$ is invariant under
the $SO(4)$ Lorentz transformations while only the $SO(3)$
symmetry is manifest at non zero $\mu$. Clearly the $SU(2)$
symmetry is also explicitly broken, at non zero chemical
potential, to $U(1)$ as for the scalar model.

For the reader's convenience we summarize the quadratic term in
the fields Lagrangian in the new variables.
\begin{eqnarray}
{\cal{L}}_{kinetic}&=&
 {1 \over 2} \psi_M \left[\delta_{MN}\Box_E-\partial_M \partial_N\right] \psi _N
+ {1 \over 2} \varphi^a_M \left[\delta_{MN}\Box_E-\partial_M
\partial_N\right] \varphi^a_N - \varphi^a_M
\varepsilon^{ab3}\left[\delta_{MN} V\cdot \partial - \frac{V_M
\partial_N + V_N
\partial_M }{2} \right] \varphi^b_N \nonumber \\
&-& \varphi^a_M \left[V\cdot V \delta_{MN}-V_M V_N
\right]\varphi^a_N
\end{eqnarray}
According to the value of the chemical potential we distinguish
two phases.

\subsubsection{The Symmetric Phase: $\mu \leq m$}
Here the $SO(4)$ Lorentz and $SU(2)$ symmetries are explicitly
broken to $SO(3)$ and $U(1)$ respectively by the presence of the
chemical potential but no condensation happens (i.e.
$<\vec{\varphi}_M>=<\psi_M>=0$). The masses at zero chemical
potential are:
\begin{eqnarray}
M^2_{\varphi_{0}^a}= M^2_{\psi_{M}}=m^2 \ , \qquad
M^2_{\varphi_{I}^a}=m^2 - \mu^2 \label{Vcurvature} \ .
\end{eqnarray}
The dispersion relations obtained by diagonalizing the 12 (4
space-time $\times$ 3 $SU(2)$ states) by 12 quadratic eigenvalue
problem leads to 3 physical vectors (i.e. each of the following
states has 3 components) with the following dispersion relations:
\begin{eqnarray}
E_{\varphi^{\mp}}=\pm\mu + \sqrt{\vec{p}^2+m^2} \ , \qquad
E_{\psi}&=&\sqrt{\vec{p}^2+m^2} \ .
\end{eqnarray}
This shows that when approaching $\mu=m$ the 3 physical components
associated with $E_{\varphi^{-}}$ will become massless signaling
an instability. Indeed we now show that for higher values of the
chemical potential a vector type condensation sets in.

\subsubsection{The Spin-Flavor Broken Phase: $\mu>M$}
In this phase the global minimum of the potential is for
$<{\varphi}_0^a>=<\psi_M>=0$, while we can choose the vev to lie
in the (Spin-Flavor) direction:
\begin{equation}
<{\varphi}_X^1>=\sqrt{\frac{\mu^2-m^2}{\lambda -
\lambda^{\prime}}} \ .
\end{equation}
We have a manifold of equivalent vacua which are obtained rotating
the chosen one under a $Z_2\times U(1)\times SO(3)$
transformation. The choice of the vacuum partially locks together
the Lorentz group and the internal symmetry while leaving unbroken
only the subgroup $Z_2\times SO(2)$. Two generators associated to
the Lorentz rotations are now spontaneously broken together with
the $U(1)$ generator.

To compute the full dispersion relations of the theory we first
need to provide the curvatures evaluated on the vacuum:
\begin{eqnarray}
\psi~~~{\rm Sector} && \nonumber \\
M^2_{\psi_0}&=&M^2_{\psi_Y}=M^2_{\psi_Z}=m^2+\lambda
\frac{\mu^2-m^2}{\lambda - \lambda^{\prime} } \ , \qquad
M^2_{\psi_X}=\mu^2 \ , \label{psi}
\\
\varphi^1~~~{\rm Sector} && \nonumber \\
\nonumber \\M^2_{\varphi_0^1}&=& \mu^2\ ,\qquad M^2_{\varphi^1_X}
=2(\mu^2 - m^2)\ , \qquad M^2_{\varphi_Y^1}= M^2_{\varphi_Z^1}=0 \ , \\
\varphi^2~~~{\rm Sector} && \nonumber \\
M^2_{\varphi_0^2}&=&m^2+\lambda \frac{\mu^2-m^2}{\lambda -
\lambda^{\prime} } \ , \qquad M^2_{\varphi_X^2}=0 \ , \qquad
M^2_{\varphi_Y^2}=M^2_{\varphi_Z^2}=\lambda^{\prime} \frac{\mu^2 -
m^2}{\lambda - \lambda^{\prime}} \ .
\end{eqnarray}
In general three states have null curvature (specifically
$M^2_{\varphi_X^2}=M^2_{\varphi_Y^1}=M^2_{\varphi_Z^1}=0$) however
for the special case $\lambda^{\prime}=0$ we have 5 zero curvature
states. To explain this behavior we not that for
$\lambda^{\prime}=0$ the potential possesses $SO(6)$ global
symmetry which breaks to $SO(5)$ when the vector field condenses.
The associated 5 states would correspond to the ordinary Goldstone
modes in the absence of an explicit Lorentz breaking. Using these
curvatures we compute the dispersion relations. {}Four of the
$\varphi$ states have (2 per each sign) the following dispersion
relations:
\begin{eqnarray}
E_{\varphi_{V}^\mp}^2&=& \vec{p}^2+\Delta^2 \mp \sqrt{4\mu^2 \,
\vec{p}^2+\Delta^4}
\end{eqnarray}
with
\begin{equation}
\Delta^2=2\mu^2 + \frac{\lambda^{\prime}}{2}\frac{(\mu^2 -
m^2)}{\lambda-\lambda^{\prime}} \ ,
\end{equation}
and to each sign we associate a vector (with two components) with respect to
$SO(2)$. In the limit of small momenta:
\begin{eqnarray}
E_{\varphi_{V}^-}^2&=&\frac{\lambda^{\prime}}{2
\Delta^2}\frac{\mu^2-m^2}{(\lambda - \lambda^{\prime})} \vec{p}^2
\nonumber +\frac{2\mu^4}{\Delta^6} |\vec{p}|^4 +
{\cal O}(p^6) = v_{\varphi^-_V}^2 \, \vec{p}^2 + \dots \ , \\
E_{\varphi_{V}^+}^2&=&2\,\Delta^2+\left[1+\frac{2\mu^2}{\Delta^2}\right]\vec{p}^2
- \frac{2\mu^4}{\Delta^6}|\vec{p}|^4 + {\cal O}(p^6) \ .
\end{eqnarray}
{}For the last two $\varphi$ physical modes (scalars of $SO(2)$)
we show directly the dispersion relations as a momentum expansion.
\begin{eqnarray}
E_{\varphi_{S}^-}^2&=& \frac{\mu^2-m^2}{3\mu^2-m^2}\vec{p}^2 +
{\cal O}(p^4) = v_{\varphi^-_S}^2 \, \vec{p}^2 + \dots \ , \qquad
E_{\varphi_{S}^+}^2=2(3\mu^2-m^2) + \gamma \vec{p}^2 + {\cal
O}(p^4) \ ,
\end{eqnarray}
where the subscripts $V,S$ stand for vector and scalar respectively, and
\begin{equation}
\gamma=\frac{(3+8c)\mu^6 +
(7-16c)m^2\mu^4+2(5c-4)m^4\mu^2-2(c-1)m^6}{\mu^2(3\mu^2-m^2)(c\mu^2-(c-1)m^2)}>0
\  ,
\end{equation}
and $c=\lambda/(\lambda - \lambda^{\prime})$. Finally for
completeness we show the three $\psi$ type state dispersion
relations:
\begin{equation}
 E_{\psi_{I}}=\sqrt{\vec{p}^2+ M^2_{\psi_I}} \ ,
\end{equation}
with $M_{\psi_I}$ defined in eq.~(\ref{psi}).
As for the scalar toy model, the velocities of our gapless modes can
be expressed directly in terms of the curvatures in the directions orthogonal
to the gapless modes as follows:
\begin{equation}
v_{\varphi^-_V}^2 = {M^2_{\varphi^2_Y} \over M^2_{\varphi^2_Y} + 4 \mu^2} \ ,
\qquad v_{\varphi^-_S}^2 = {M^2_{\varphi^1_X} \over M^2_{\varphi^1_X} + 4 \mu^2}
\ .
\end{equation}
We find that the number of gapless states emerging when the
flavor$\times$rotational symmetry spontaneously breaks is
insensitive to the choice of the coefficients in the effective
potential. However the momentum dependence depends crucially on
the symmetries of the potential. The full vector Lagrangian (i.e.
kinetic $-$ potential) possesses the $Z_2\times U(1)\times SO(3)$
(flavor$\times$rotational) symmetry which breaks, at non zero
chemical potential, when the vector condenses to $Z_2\times
SO(2)$. The gapless modes are associated to the 3 broken
generators. One state is a scalar of $SO(2)$ while the other two
fields constitute a vector of $SO(2)$. The scalar Goldstone
possesses linear dispersion relations independently of the choice
of the vector potential. This is the Goldstone associated to the
$U(1)$ breaking of the previously investigated scalar toy model.
However for the dispersion relations of the $SO(2)$ vector state
we have either quadratic or linear dispersion relations. The
quadratic ones occur when the potential parameters are such that
it possesses an enhanced $SO(6)$ symmetry. When condensing the
vector breaks the potential symmetries $SO(6)$ to $SO(5)$ while
the kinetic term (i.e. the derivative part of the Lagrangian) is
still invariant under the continuous $U(1)\times SO(3)$ only. In
this case the potential has 5 flat directions (i.e. 5 null
curvatures (occurring for $\lambda^{\prime}=0$)) and we would
count in the spectrum 2-massless vectors and a scalar of $SO(2)$.
However the reduced symmetry of the kinetic term prevents the
emergence of another gapless vector type mode while turning the
linear dispersion relations of one of the vector states into
quadratic ones. As for the toy model case, at the second order
phase transition point, all the velocities of the gapless modes
vanish regardless of the value of the coupling constants. Again
the information of the symmetries (conformal and/or global) of the
potential, at nonzero chemical potential, are nicely transferred
from the curvature  to the momentum dependence of the dispersion
relations via the velocities of the gapless modes. Our result is
in agreement with the Chadha-Nielsen counting scheme.

\section{Conclusions and Physical Applications}
\label{applications}

We summarized our findings in the Table I and conclude our work by
considering some possible physical applications of massive
relativistic vector condensation at high chemical potential.
\subsection{2 Color QCD}

Two color QCD at non zero chemical potential recently attracted a
great flurry of interest since lattice simulations can be reliably
performed at non zero quark chemical potential.  In
\cite{Lenaghan:2001sd} vector condensation at non zero quark
chemical potential has been predicted for two color QCD with an
even numbers of flavors. The predictions are supported
\cite{MPLombardo} by lattice studies. Here we confirmed the
previous results and provided a more detailed study of the vector
condensation phenomenon at non zero chemical potential.

An interesting direct application of our results is for QCD with
two colors and one light flavor. This theory has global quantum
symmetry group $SU(2)$. The extra classical axial $U_A(1)$
symmetry is broken by the Adler-Bell-Jackiw anomaly.  At zero
chemical potential Lorentz invariance cannot be broken and the
simplest bilinear condensate (see Ref.~\cite{Lenaghan:2001sd} for
conventions) is of the type:
\begin{equation}
\epsilon^{c_1 c_2} \epsilon^{\alpha \beta}
Q_{\alpha,c_1}^{I}Q_{\beta,c_2}^{J} E_{IJ} \ ,
\end{equation}
with $E=2iT^2$ the antisymmetric matrix in the flavor space and
$Q_{\alpha c}^I$ is a Weyl spinor with $\alpha=1,2$ the spinorial
index, $c=1,2$ the color and $I=1,2$ the flavor indices. The
generators of $SU(2)$ in the fundamental representation are
$T^a=\tau^a/2$ with $\tau^a$ the standard Pauli matrices. It is
easy to check that $SU(2)$ remains unbroken since we have
${T^a}^T\, E+E\,T^a=0$ for all of the $SU(2)$ generators. So we
have no Goldstone bosons at all in the theory. However we do have
physical vectors (i.e. spin 1 fields) and massive scalar states at
zero density. These states should be comparable in mass (recall
the $\eta^{\prime}$ versus the octet of ordinary vector bosons in
3 color QCD). The massive spin one fields (i.e. the hadrons) must
transform according to the regular representation of $SU(2)$ and
carry non trivial baryon number which is proportional to $T^3$.

Our analysis can be immediately used to predict that when turning
on a non zero baryon chemical potential a superfluid phase
transition will set in for $\mu$ larger then the mass of the
vector fields. This is so since we always have at least one
gapless mode with linear dispersion relations. We also predict the
spontaneous breaking of rotational invariance and the emergence of
a vector state (i.e. doublet of $SO(2)$) with either linear or
quadratic dispersion relations. It is straightforward to extend
and apply our analysis to the case of different number of flavors
and to ordinary QCD at finite isospin chemical potential.
\cite{Lenaghan:2001sd}.

\subsection{Extra-dimensions}

Finally we extend our model to an $SU(2)$ charged vector in $D-1$
number of space dimensions (with $SO(D)$ Euclidean Lorentz
symmetry) with an associated non zero chemical potential. In this
case the chemical potential first breaks explicitly $SO(D)$ to the
spatial $SO(D-1)$. At sufficiently high chemical potential
$SO(D-1)$ breaks spontaneously to $SO(D-2)$. The gapless states
consist of a scalar field under the $SO(D-2)$ transformations with
linear dispersion relations and one vector of $SO(D-2)$ with
either linear or quadratic dispersion relations.
\begin{table}
\begin{tabular}{ccc}
~&$\lambda > \lambda'$ & $\lambda' = 0$\\
\hline
Lagrangian & $Z_2\times U(1)\times SO(3)\longrightarrow
Z_2\times SO(2)$ & $Z_2\times U(1)\times SO(3)\longrightarrow
Z_2\times SO(2)$ \\
Potential & $Z_2\times U(1) \times SO(3)\longrightarrow Z_2\times
SO(2)$ & $Z_2\times SO(6)\longrightarrow Z_2\times SO(5)
$ \\
\hline
\# of broken generators: &  &  \\
 Lagrangian  & 3 & 3 \\
 Potential   & 3 & 5 \\
\hline
spectrum of Goldstone modes: &  &  \\
Type I ($E\propto |\vec{p}|$) &  1~SO(2) Scalar + 1~SO(2) Vector & 1~SO(2) Scalar  \\                                \\
Type II ($E\propto |\vec{p}|^2$) & $-$ &  1~SO(2) Vector  \\
\end{tabular}
\caption{Summary of the relevant symmetries and spectrum of the
Goldstone modes for the parameter choices $\lambda > \lambda'$ and
$\lambda'=0$ of the fourth--order potential. We also indicate in
the first row the symmetries (after the arrows) which are not
broken by the vector condensate.}
\end{table}

\acknowledgments
It is a pleasure for us to thank M.P. Lombardo, H.B. Nielsen, and
P. Olesen for helpful discussions. We are also indebt to J.
Schechter for discussions and careful reading of our manuscript.
The work of F.S. is supported by the Marie--Curie fellowship, while
W.S. acknowledges support by DAAD.

\end{document}